\documentclass[preprint,12pt]{elsarticle}

\usepackage{graphicx}

\usepackage{amssymb}

\usepackage{amsmath}
\newtheorem{theorem}{Theorem}

\journal{Journal Name}

\begin{document}
	
	\begin{frontmatter}
		
		

		\title{Analytical Factors for Describing the Control Ability of Linear Discrete-time Systems 
			\thanks{Work supported by the National Natural Science Foundation of China (Grant No. 61273005)}}
		
		
		\author{Mingwang Zhao}
		
		\address{Information Science and Engineering School, Wuhan University of Science and Technology, Wuhan, Hubei, 430081, China \\
			Tel.: +86-27-68863897 \\
			Work supported by the National Natural Science Foundation of China (Grant No. 61273005)}
		
		\begin{abstract}
			 In this paper, the analytical volume computations of the zonotopes generated by the matrix pair with $n$ different or repeated real eigenvalues are discussed firstly, and then by deconstructing the volume computing equations, 3 classes of the shape factors are constructed. These analytical volume and shape factors can describe accurately the size and shape of the zonotopes.
			Because the control ability for LDT systems with the unit input variables (i.e. the input variable is with bounded value as 1) is directly related to the controllability region \cite{zhaomw202003}, based on these analytical expressions on the volume and shape factors, the control ability can be quantized conveniently. By choosing these analytical expressions as the objective function or constraint conditions, a novel optimizing problems and solving method for the control ability can be founded. Based on the optimization, not only the open-loop control ability, and but also the some closed-loop control performance, such as the optimal time waste, robustness of the control strategy, etc, can be promoted, according to the conclusions in paper \cite{zhaomw202003}

		\end{abstract}
		
		
	\end{frontmatter}
	
	
	\section{Introduction}
	\label{S:1}	
	
	The controllability analysis for the dynamical systems is started in 1960's by R. Kalman, et al, \cite{KalmHoNar1963}, and became one of the most important analysis methods for dynamical properties and supported the 60 years development of the control theory, for revealing deeply the possibility controlling the state variable by the input variables.
	
	The classical controllability concept is a qualitative concept with two-value logic and the dynamical systems is distinguished as only two classes of systems, controllable systems and uncontrollable systems. The concept and corresponding criterion could not tell us the control ability and control efficiency of the input variable to the state variables. In fact, the quantitative concept and analysis method are very important for the control theory and engineering, and many engineering problems are dying to these concept and method. 
	
	In past 60 years, some works stated as follows are on the quantitation of control ability.
	
	1) Based the definition of the controllability Grammian matrix $G_{N}$ and the controllability Ellipsoid $E_{N}$, in papers \cite{VanCari1982}, \cite{Georges1995}, \cite{PasaZamEul2014}, and \cite{Ilkturk2015}, the determinant value $ \det \left( G_{N} \right) $ and the minimum eigenvalue $ \lambda _{\textnormal {min} } \left (G_{N} \right ) $ of the controllability Grammian matrix $G_{N}$, correspondingly the volume $ \textnormal {vol} \left( E_{N} \right)$ and the minimum radius $ r _{\textnormal {min}} \left (E_{N} \right ) $ of the controllability ellipsoid $E_{N}$, can be used to quantify the control ability of the input variable to the state space, and then be chosen as the objective function for optimizing and promoting the control ability of the linear dynamical systems. Due to lack of the analytical computing of the determinant $ \det \left( G_{N} \right) $ and eigenvalue $ \lambda _{\textnormal {min}} \left (G_{N} \right ) $, correspondingly the volume $\textnormal {vol} \left( E_{N} \right)$ and the radius $ r _{\textnormal {min}} \left (E_{N} \right ) $, these optimizing problems for the control ability are solved very difficulty, and few achievements about that have been made.
	
	2) Based the definition of the controllability region $R_{N}$, the minimum distance of the boundary of the region 
	$R_{N}$ is used to measure and optimize the control ability \cite{ViswLongLik1984}. Because of the difficulty to compute the distance, the further works have not been reported.
	
	3) Based on the well-known PBH test for the controllability, a degree of modal controllability was put forward to measure and optimize the control ability \cite{LonSirSev1982} \cite{HamElad1988} \cite{
		HamJard2014}. Because the degree is only for the single modal and not for the whole system, the further works have not been carried forward. 
	
	The control ability for the linear discrete-time (LDT) systems is defined and a relation theorem among the open-loop control ability, the control strategy space (i.e., the solution space of the input variables for the control problems), and the some closed-loop control performance, such as, the optimal time waste, the response speed, the robustness of the the control strategy, etc, is purposed and proven in paper \cite{zhaomw202003}. By the theorem in paper \cite{zhaomw202003}, we can see, optimizing the control ability of the open-loop systems is with the great significations for promoting the performance indices of the closed-loop control systems. 
	
	Paper \cite{zhaomw202001} puts forward an analytical computing equation method for the volume of the zonotope generated by the matrix pair $(A,B)$ with $n$ different eigenvalues of matrix $A$. Because the controllability region definied in papers \cite{Hu2002} \cite{HuMiQiu2002}, is indeed a zonotope discussed in paper \cite{zhaomw202001}, based on the results in papers \cite{zhaomw202001} and \cite{zhaomw202003}, studies on measuring and optimizing for the control ability can be carried out. In this paper,  the analytic volume computation of the zonotopes generated by the matrix pair $(A,B)$ with the repeated eigenvalues of matrix $A$
	 is discussed firstly, and then some analytical factors described the size and shape of the zonotope, that is, the measurement for the control ability for linear systems, can be founded for analyzing and optimizing the control ability. These analytic computing methods and results proposed in this paper will be became the basis for analysis and optimization of the control ability of the LDT systems.

	\section{Analytic factors for Control ability of the linear systems with $n$ different Eigenvalues}
	
	\subsection {Definition of the Control Region}
	
	In this paper, the quantitative analysis of the control ability will be carried out for the following linear discrete-time (LDT) systems 
	\begin{equation}
	x_{k+1}=Ax_{k}+Bu_{k}, \quad x_{k} \in R^{n},u_{k} \in R^{r}, \label{eq:a01}
	\end{equation}
	\noindent where $x_{k}$ and $u_{k}$ are the state variable and input
	variable, respectively, and matrices $A \in R^{n \times n}$ and $B \in
	R^{n \times r}$ are the state matrix and input matrix, respectively,
	in the system models \cite{Kailath1980}, \cite{Chen1998}. 
	
	the so-called control ability of the dynamical systems is the ability to control one state in the state space to the other state by the input variables. In control theory, the control ability can be divided into two cases, one is for controlling the initial state to the original of the state space and other one is for controlling the initial state in the original to the expecting state. The two classes of control ability analysis are called as state controllability and reachability, respectively. In fact, with the help with of the solution expression of the state equation \eqref{eq:a01},  the analysis of state controllability and reachability for the systems \eqref{eq:a01} are carried out on the following equations, respectively
	\begin{align}
	0 & = A^N x_0+ \sum _{k=0}^{N-1} A^{N-k-1}Bu_k  ,\; \forall x_0\in R^n \\
	x_1 & = \sum _{k=0}^{N-1} A^{N-k-1}Bu_k,\; \forall x_1\in R^n
	\end{align}
	that is, 
	\begin{align}
	& P^c_N (A) \times u_{0,N-1} =-x_0 ,\; \forall x_0\in R^n \\
	& P^d_N (A) \times u_{N-1,0} =x_1 ,\; \forall x_1\in R^n
	\end{align}
	where 
	$P^c_N (A) =\left[ A^{-1}B,A^{-2}B,\dots, A^{-N}B \right]$, 
	$P^d_N (A) =\left[B,AB,\dots, A^{N-1}B \right]$,
	$u_{0,N-1} = \left[u_0^T,u_1^T,\dots,u_{N-1}^T \right] ^T$,
	$u_{N-1,0} = \left[u_{N-1}^T,u_{N-2}^T,\dots, u_0^T \right] ^T$,
	
	The input variables of the most Engineering controlled plants are with bounds, that is, the input variables are subjected to certain constraints. By the normalization of these bounds of the input variable, these constraints can be noted as follows
	\begin{align}
	\Vert u_k \Vert \le 1 \label{eq:as02}
	\end{align}
	In fact, for the systems with unbounded input variable, the constraints \eqref{eq:as02} can be described for the unit input variables. Whether the input variables are bounded or unbounded, the controllability region for the two classes of control problems are defined respectively as follows \cite{Hu2002}, \cite{HuMiQiu2002},\cite{zhaomw202003}
	\begin{align}
	& R^c_N =\left\{ x:x= P^c_N(A) \times U,\; \forall U\in R^{n\times N} \right\} \label{eq:as005} \\
	& R^d_N =\left\{ x:x= P^d_N(A) \times U,\; \forall U\in R^{n\times N} \right\} \label{eq:as006}
	\end{align}
	Because that $P^c_N(A)=A^{-1} P^d_N \left(A^{-1} \right) $, the two classes of the controllability region can be transformed each other. Therefore, the control ability based on the region $R^d_N$ are discussed later, and the obtained results can be generalized conveniently to the region $R^c_N$. 
	
	\subsection {Analytic Factors for Describing the Control Ability}
	
	As pointing out in paper \cite{zhaomw202001}, \cite{zhaomw202003}, the size and shape of the controllability region $R^{c}_{N}$ has direct relations to the control ability. The relevant conclusions can be summarized in the following theorem \cite{zhaomw202003}.
	
	\begin{theorem} \label{la:ab02} 
		It is assumed that two LDT Systems $\Sigma_1$ and $\Sigma_2$ are controllable, and their controllability regions are $R^{c}_{N,1} $ and $R^{c}_{N,2} $ respectively.
		If we have
		\begin{align} \label{eq:abc01}
		R^{c}_{i,1} (i) \subseteq R^{c}_{i,2} (i),\; \forall i\le N,
		\end{align}
		for the control problem stabilizing the state $x_0 \left (x_0 \in R^d_{N,1} \cap R ^{c}_{N,2} \right) $ to the original of the state space,
		the following conclusions hold under the input amplitude constraint.
		
		1) The time waste of the time-optimal control for the system $\Sigma_2$ is not more than that of $\Sigma_1$, that is, there exist some control strategies with the less control time and the faster response speed for the system $\Sigma_2$. 
		
		2) There exist more control strategies for the system $\Sigma_2$, that is, 
		the bigger the controllability region is, the bigger the solution space of the input variable for the control problems, the easier designing and implementing the control are. 
	\end{theorem}
	
	In a word, according to the above theorem, the bigger the controllability region $R^{c}_{N} $ is, the stronger the control ability is, and then the better the closed-loop performances related to the waste time of the control process
	are. The results in \textbf{Theorem \ref{la:ab02}} can be generalized conveniently to the reachablity region $R^{d}_{N} $. Therefore, based on the theorem, quantifying and maximizing the controllability region $R^{c}_{N}$ and reachability region $R^{d}_{N}$ are indeed for quantifying and maximizing the control ability. Out of the need of the practical control engineering problem, it is highly necessary to establish the quantifying and maximizing method about $R^{c}_{N}$ and $R^{d}_{N}$.
	
	In paper \cite{zhaomw202001}, an analytic volume-computing equation for the zonotope generated by the matrix pair $(A,B)$ with $n$ different real eigenvalues of matrix $A$ is proven. In fact, the reachability region $R^{d}_{N}$ defined by Eq. \eqref{eq:as005}
	is a special zonotope defined in paper \cite{zhaomw202001}. Therefore, the quantifying and maximizing of the reachability region $R^{d}_{N}$ can be carried out based on the results in paper \cite{zhaomw202001}. The results in the volume computation of the zonotope generated by the matrix pair $(A,B)$, in \textbf{ Theorem 2} and \textbf{ Theorem 3} of paper \cite{zhaomw202001} can be summarized as the following theorem.
	
	\begin{theorem} \label {t30} If matrix $A\in R^{n\times n}$ is with $n$ different eigenvalues in the interval $[0,1)$ and $b \in R^n$ is a vector, the volume of the infinite-time zonotope $R_\infty$ generated by matrix pair $ ( A, B) $ can be computed analytically by the following equation:
		\begin{equation} \label{eq:as17}
		\textnormal {vol} (R_\infty) = \left| \det(P) \left( \prod_{1 \leq j_{1}<j_{2} \leq n} \frac{ \lambda_{j_{2}}- \lambda_{j_{1}}}{1- \lambda_{j_{1}} \lambda_{j_{2}}} \right) \left( \prod_{i=1}^{n} \frac{q_{i} b}{1- \lambda_{i}} \right) \right| 
		\end{equation}
		where $\lambda_{i}$ and $q_{i}$ are the $i$-th eigenvalue and the corresponding unit left eigenvector of matrix $A$, matrix $P$ is the matrix transforming the matrix $A$ as a diagonal matrix.
	\end{theorem} 
	
	Based on the above theorem, the volume of the reachability region $R^d_N$ when $N \rightarrow \infty$ can be computed analytically.
	Furthermore, some analytical factors describing the shape of the $R^d_N$ can be got by deconstructing the analytical volume computing equation \eqref{eq:as17}. These analytical expressions for these volume and shape factors can be describe quantitatively the control ability of the dynamical systems, and then optimizing these volume and shape factors is indeed maximizing the control ability. 
	
	\subsection{Decoding the Controllability Ellipsoid}
	
	According to the volume computing equation \eqref{eq:as17}, some factors described the shape and size of the reachability region, that is, the control ability of the dynamical systems, are deconstructed as follows.
	\begin{align} 
	F_1 & = \left \vert \prod_{1 \leq j_{1}<j_{2} \leq n} \frac{ \lambda_{j_{2}}- \lambda_{j_{1}} }{ 1- \lambda_{j_{1}} \lambda_{j_{2}}} \right \vert \label{eq:630} \\
	F_{2,i} 
	&= \frac{ \left \vert q_ib \right \vert}{ 1- \lambda_{i} } 
	\label{eq:640} \\
	F_{3,i} & =\left \vert q_ib \right \vert \label{eq:650}
	\end{align}
	The above analytical factors can be called respectively as 
	the shape factor, the side length of the circumscribed rhombohedral, and the modal controllability. In fact, the shape factor $F_1$ is also the eigenvalue evenness factor of the linear system, and can describe the control ability caused by the eigenvalue distribution. In addition, the modal controllability factor $F_{3,i}$ have been put forth by papers \cite{chan1984} \cite{HamElad1988} \cite{ChoParLee2000} \cite{Chenliu2001}, and will not be discussed here.
	
	\subsection {The Shape Factor of the Reachability Region and the Eigenvalue Evenness Factor of the Linear System}
	
\begin{figure}[htbp]
		\centering
		\begin{minipage}[c]{0.45\textwidth} 
			\centering
			\includegraphics[width=0.8\textwidth]{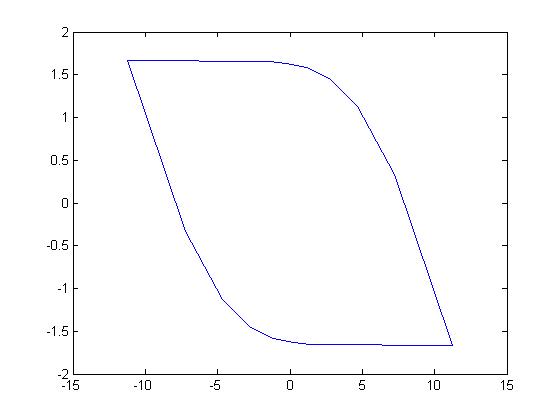} 
			\\ (a) \footnotesize {(0.4,0.9,0.7813)}
		\end{minipage}%
		\begin{minipage}[c]{0.45\textwidth}
			\centering
			\includegraphics[width=0.8\textwidth]{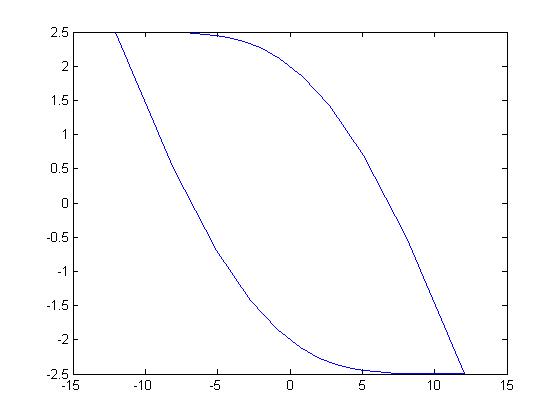}
			\\ (b) \footnotesize {(0.6,0.9,0.6522)}
		\end{minipage}
		\begin{minipage}[c]{0.45\textwidth}
			\centering
			\includegraphics[width=0.8\textwidth]{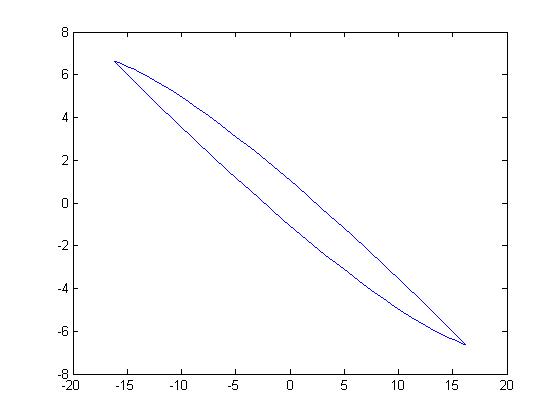}
			\\ (c) \footnotesize {(0.85,0.9,0.2128)}
		\end{minipage}
		\caption[c]{The 2-dimensional zonotope with $(\lambda_1,\lambda_2,F_1)$ \label {fig:as01} }	
	\end{figure}

\begin{figure}[htbp]
		\centering
		\begin{minipage}[c]{0.45\textwidth} 
			\centering
			\includegraphics[width=0.8\textwidth]{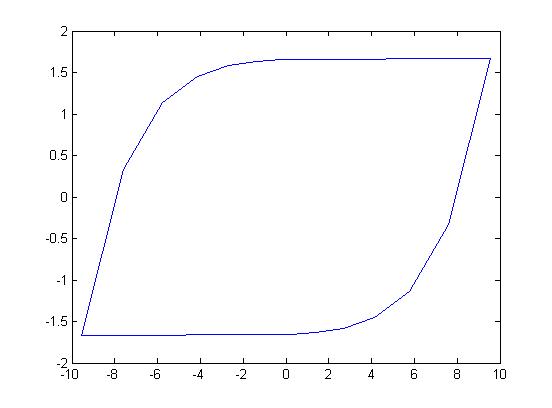} 
\\ (a) \footnotesize {(0.4,0.9,0.7813)}
		\end{minipage}%
		\begin{minipage}[c]{0.45\textwidth}
			\centering
			\includegraphics[width=0.8\textwidth]{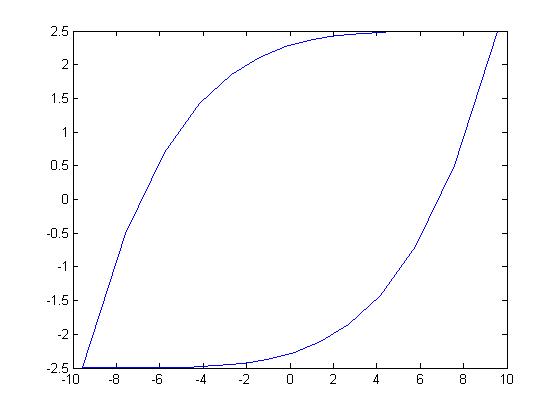}
\\ (b) \footnotesize {(0.6,0.9,0.6522)}
		\end{minipage}
		\begin{minipage}[c]{0.45\textwidth}
			\centering
			\includegraphics[width=0.8\textwidth]{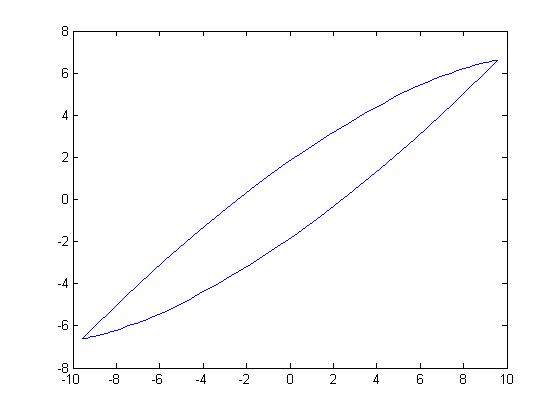}
\\ (c) \footnotesize {(0.85,0.9,0.2128)}
		\end{minipage}
	\caption[c]{The 2-dimensional zonotope with $(\lambda_1,\lambda_2,F_1)$ \label {fig:as02} }	
	\end{figure}

	Fig. \ref{fig:as01} shows the 2-dimensional zonotopes $R^d_{30} $, i.e., the sampling number $ N=30$, generated by the 3 matrix pairs $(A,b)$ that the matrix $A$ is with the different eigenvalues and matrix $b$ is a same vector, and Fig. \ref{fig:as02} shows the 2-dimensional zonotopes generated by the diagonal matrix pairs of these 3 matrix pairs $(A,b)$, that is, the zonotopes in Fig. \ref{fig:as02} are in the invariant eigen-space.

 From these figures, we can see, when the two eigenvalues of the matrix $A$ are approximately equal, the minimum distances of the boundary of the region $R_N^d$ to the original  will be approximately zero, and the region $R_N^d$ will be flattened. Similar casees are also for the n-dimensional zonotope generated by the matrix pair.
Therefore, the distributions of all eigenvalues of the matrix $A$ are even, the ratio between the minimum and maximum distance of the boundary of the zonotope generated by the pair $(A,b)$ can be avoided as a small value and the zonotope will be avoided flattened. 
		
	The factor $F_1$ deconstructed from the volume computing equation \eqref{eq:as17} can be used to describe the uniformity of the $n$ biggest distances of the region $ R_N^d$ in $n$ eigenvectors. The bigger the value of the factor $F_1$, the bigger 
	the ratio between the minimum and maximum distances of the bounded of the region $\overline R_N^d$ is, and then the greater the volume of the region is.
	
	Otherwise, the factor $F_1$ can be used to describe the evenness of the eigenvalue distribution of the linear system $\Sigma(A,B)$. The bigger the value of the factor $F_1$, the bigger the controllable region of the system is, and the stronger the control ability of the systems is.
	
	\subsection {The circumscribed hypercube and circumscribed rhombohedral of the reachability region}
	
	The factor $F_{2,i}$ is indeed the biggest distance of the region $ R_N^d$ in the $i$-dimensional coordinate of the eigen-space (shown as in  Fig. \ref{fig:as02} ), that is, the $n$ side lengths of the circumscribed hypercube of the region $R_N^d$ are $2F_{2,i},i=\overline{1,n}$. By the volume equation \eqref{eq:as17}, the volume of the region can be represented as the production of the volume $\prod _{i=1} ^{n} F_{2,i}$ of the circumscribed hypercube and the shape factor $F_1$
	
	Because that these expressions of the volume and the shape factors can describe accurately the size and shape of the zonotope generated by the matrix pair, i.e., the control ability of the dynamical systems \cite{zhaomw202003}, these expressions can be used conveniently to be the objective function or the constrained conditions for the optimizing problems of the control ability of dynamical systems.

	\section{The Control Ability for the Systems with the Repeated real Eigenvalues}
	
	\subsection{A Property about the Control Ability for the Systems with the Repeated Eigenvalues}
	
	In the last section, the volume of the zonotope generated by the matrix pair $(A,b)$, correspondingly the control ability of the linear systems, is discussed for the matrix $A$ with $n$ different real eigenvalues. Next, the volume is discussed for that the matrix $A$ is with the repeated real eigenvalues and the matrix $B$. 
	
	When some eigenvalues of matrix $A$ are the repeated eigenvalues, the matrix $A$ can be transformed as a Jordan matrix by a similar transformation. In view of this, the following discussion are only carried out for the Jordan matrix. For the volume of the zonotope generated by the Jordan matrix pair $(A,b)$, a following theorem about the relation between the volume and matrix $b$ can be stated and proven firstly.
	
	\begin{theorem} \label{th3201}
		The volume of the zonotope generated by the Jordan matrix pair $(A,b)$ has relation only to the last rows of the matrix blocks of the matrix $b$, corresponding to the each Jordan block of the matrix $A$, and has no relation to other rows.
	\end{theorem}
	
	The conclusion in \textbf{ Theorem \ref{th3201}} is consistent with the conlusion, in control theory, that 
	the state controllability of the systems $ \Sigma(A,b)$ has relation only to the last rows of these matrix blocks of the matrix $b$, corresponding to the each Jordan block of the Jordan matrix $A$, and has no relation to other rows. According to \textbf{ Theorem \ref{th3201}}, for simplifying the volume computation for the Jordan matrix pair $(A,b)$, the rows of matrix $b$ entirely unrelated to the volume will be regarded as 0.
	
	\textbf{ Proof:} Without loss of the generality, the theorem is proven only for the Jordan matrix with one Jordan block, and other cases can be proven similarly.
	
	When the matrix $A$ is with only one Jordan block, the two matrix of the matrix pair can be written as follows
	\begin{equation} \label{eq:c3b01}
	A= \left[ \begin{array}{ccccc}
	\lambda & 1 & 0 & \cdots & 0\\
	0 & \lambda & 1 & \cdots & 0\\
	0 & 0 & \lambda & \cdots & 0\\
	\vdots & \vdots & \vdots & \ddots & \vdots\\
	0 & 0 & 0 & \cdots & \lambda
	\end{array} \right], \quad b= \left[ \begin{array}{c}
	b_{1}\\
	b_{2}\\
	b_{3}\\
	\vdots\\
	b_{n}
	\end{array} \right]
	\end{equation}
	\noindent Noting 
	\begin{equation}
	\zeta_{n,k}=A^{k}b= \left[ \begin{array}{c}
	\sum_{i=1}^{n}C_{k}^{i-1} \lambda^{k-i+1}b_{n-i}\\
	\vdots\\
	\frac{k(k-1)}{2} \lambda^{k-2}b_{n}+k \lambda^{k-1}b_{n-1}+ \lambda^{k}b_{n-2}\\
	k \lambda^{k-1}b_{n}+ \lambda^{k}b_{n-1}\\
	\lambda^{k}b_{n}
	\end{array} \right]
	\end{equation}
	\noindent where $C_{k}^{i-1}= \frac{k!}{(i-1)!(k-i+1)!}$ is the binomial coefficient. when $k-i+1<0$, noting
	\begin{equation}
	C_{k}^{i-1} \lambda^{k-i+1}=0
	\end{equation}
	\noindent If $b_n \ne 0$, it is can be proven that the transformation matrix
	\begin{equation} 
	P= \left[ \begin{array}{ccccc}
	1 & - \frac{b_{n-1}}{b_{n}} & \cdots & - \frac{b_{2}}{b_{n}} & - \frac{b_{1}}{b_{1n}}\\
	& \ddots & \ddots & \vdots & \vdots \\
	& & 1 & - \frac{b_{n-1}}{b_{n}} & - \frac{b_{n-2}}{b_{n}} \\
	& & & 1 & - \frac{b_{n-1}}{b_{n}}\\
	& & & & 1
	\end{array} \right]
	\end{equation}
	\noindent can make the followings equation hold
	\begin{equation}
	P \zeta_{n,k}= \xi_{n,k}=b_{n} \left[ \begin{array}{c}
	C_{k}^{n-1} \lambda^{k-n+1}	\\
	\vdots \\
	\frac{k(k-1)}{2} \lambda^{k-2}\\
	k \lambda^{k-1}\\
	\lambda^{k}
	\end{array} \right]
	\end{equation}
	
	For the non-zero eigenvalue $ \lambda$, there exists an elementary transformation matrix $Q$ to make the following equation holds.
	\begin{equation}
	QP \zeta_{n,k}=Q \xi_{n,k}= \beta_{n,k}=b_{n} \left[ \begin{array}{c}
	\frac{k^{n-1}}{(n-1)!} \lambda^{k-n+1} \\
	\vdots \\
	\frac{k^2}{2} \lambda^{k-2}\\
	k \lambda^{k-1}\\
	\lambda^{k}
	\end{array} \right]
	\end{equation}
	\noindent that is, $ \xi_{n,k}$ and $ \beta_{n,k}$ have relations only to $b_{n}$ and have not relations to $b_i (i\ne n)$. In addition, from the expressions of the transformation matrices $P$ and $Q$, we have
	$$ \det(P)=1 \; \textnormal {and} \; \det(Q)=1$$
	Therefore, By the volume computing equation (6) in paper \cite{zhaomw202001} the volume of the zonotope generated by the above Jordan matrix pair $(A,B)$ is
	\begin{align}
	\textnormal{Vol}(C_{z,N})	& =V_{n} \left(C_{n} \left([B,AB,...,A^{N-1}B] \right) \right) \notag \\
	&=V_{n} \left(C_{n} \left([ \zeta_{n,0}, \zeta_{n,1},..., \zeta_{n,N-1}] \right) \right) \notag \\
	&= \sum_{(k_{1},k_{2}, \cdots,k_{n}) \in \Omega_{0,N-1}^{n}} \left| \mathrm{det} \left( \left[ \zeta_{n,k_{1}}, \zeta_{n,k_{2}}, \cdots, \zeta_{n,k_{n}} \right] \right) \right| \notag \\
	&= \left| \mathrm{det} \left(P^{-1}Q^{-1} \right) \right| \sum_{(k_{1},k_{2}, \cdots,k_{n}) \in \Omega_{0,N-1}^{n}} \left| \mathrm{det} \left( \left[ \beta_{n,k_{1}}, \beta_{n,k_{2}}, \cdots, \beta_{n,k_{n}} \right] \right) \right| \notag \\
	&= \sum_{(k_{1},k_{2}, \cdots,k_{n}) \in \Omega_{0,N-1}^{n}} \left| \mathrm{det} \left( \left[ \beta_{n,k_{1}}, \beta_{n,k_{2}}, \cdots, \beta_{n,k_{n}} \right] \right) \right|
	\end{align}
	
	Because that $ \beta_{n,k_{i}}$ in the last equation has relation only to $b_{n}$, the volume of the zonotope is with relation only to the last row of the matrix $b$ in the Jordan matrix pair $(A,b)$, and isn't with relation to the other rows.
	
	Therefore, according to the theorem, the volume computation of the zonotope for the Jordan matrix pair $(A,b)$ can be equivalent to the volume computation for the Jordan matrix pair $(A, \beta)$, where $ \beta=[0, \dots,0, b_n]^T$.
	
	\subsection{The volume computation of the infinite-time zonotope for the Jordan matrix $A$ only with a Jordan block }
	
	Nest, the volume computation of the infinite-time zonotope for the Jordan matrix $A$ is only with a Jordan block is discussed and the obtained results are can be generalized to the other cases.
	
	When all eigenvalues of the matrix $A$ satisfy $ \lambda \in [0,1) $, the infinite-time zonotope generated by the pair, that is, $N \rightarrow \infty $, is a finite geometry, and its volume is also a finite value. For the analytic computing of the volume of the infinite-time zonotope generated by the Jordan matrix pair \eqref{eq:c3b01}, we have the following theorem. 
	
	\begin{theorem} \label{c3bt01} When all eigenvalues of the matrix $A$ satisfy $ \lambda \in [0,1) $, the volume of the infinite-time zonotope generated by the Jordan matrix pair with one-Jordan-block matrix $A$ is computed as follows 
		\begin{equation} \label{eq:e3d019}
		V = \frac{b_{n}^{n}}{ \left(1- \lambda \right)^{n} \left(1- \lambda ^{2} \right)^{n(n-1)/2}}
		\end{equation}
	\end{theorem}
	
	\textbf{Proof}: Next, the theorem is proven by the approximation method. For that, a matrix $A$ with $n$ different eigenvalues can be designed as follows
	\begin{equation} \label{eq:e3d020}
	A= \left[ \begin{array}{ccccc}
	\lambda & 1\\
	& \lambda+ \varDelta & 1\\
	& & \lambda+2 \varDelta & \ddots\\
	& & & \ddots & 1\\
	& & & & \lambda+(n-1) \varDelta
	\end{array} \right]
	\end{equation}
	where $ \varDelta >0 $ is a sufficient small positive number, the eigenvalues $ \left \{ \lambda_{i}= \lambda+(i-1) \varDelta,i= \overline{1,n} \right \} $ are different and are also in the interval $[0,1)$. In fact, when $ \varDelta \rightarrow 0 $, the matrix $A$ is also a one-Jordan-block matrix, and then the volume of the zonotope generated by the matrix pair $(A,b) $ is that of the zonotope generated by the Jordan matrix pair.
	
	In addition, for the volume computing of the zonotope, by \textbf{ Theorem \ref{c3bt01}}, matrix $b$ can be regarded as $b=[0, \dots,0,b_n]^T$. 
	Therefore, for all $ \lambda \in [0,1)$, Eq. \eqref{eq:e3d019} can be proven as follows.
	
	(1) According to the knowledge of matrix theory, the matrix for transforming the matrix $A$ in Eq. \eqref{eq:e3d020} as a diagonal matrix can be constructed as follows
	\begin{equation}
	P= \left[p_{ij} \right]_{n \times n}= \left[ \begin{array}{ccccc}
	1 & 1 & 1 & \cdots & 1\\
	0 & \varDelta & 2 \varDelta & \cdots & (n-1) \varDelta\\
	0 & 0 & 2 \varDelta^{2} & \cdots & (n-1)(n-2) \varDelta^{2}\\
	\vdots & \vdots & \vdots & \ddots & \ddots\\
	0 & 0 & 0 & \cdots & (n-1)! \varDelta^{n-1}
	\end{array} \right]
	\end{equation}
	where the $n$ columns of the transformation matrix $P$ is the $n$ eigenvectors of matrix $A$ with $n$ different eigenvalues,
	\begin{align}
	p_{ij} &	= \left \{ \begin{array}{ll}
	0, & i>j\\
	1, & i=1\\
	(j-1)...(j-i+1) \varDelta^{i-1}= \frac {(j-1)!} {(j-i)!} \varDelta^{i-1}, & \textrm{others}
	\end{array} \right. \\
	\det \left(P \right) & = \prod_{i=1}^{n-1} \left(i! \varDelta^{i} \right)= \left( \prod_{i=1}^{n-1}i! \right) \varDelta^{n(n-1)/2}
	\end{align}

	And then, the corresponding diagonal matrix pair obtained by transformation is as
	\begin{align}
	\overline{A} & = \textrm{diag} \left \{ \lambda_{1}, \lambda_{2}, \cdots, \lambda_{n} \right \} 
	\\
	\overline{b} &=P^{-1}b= \left[ \beta_{1}, \beta_{2}, \cdots, \beta_{n} \right]^{T}
	\end{align}
	In fact, the matrix $ \overline{b}$ can be regarded as the solution of the 
	triangular matrix equation $P \overline{b}=b$, that is, 
	\begin{equation} \label{eq:e3d024}
	\left[ \begin{array}{cccccc}
	p_{11} & p_{12} & \cdots & p_{1,n-1} & p_{1,n}\\
	0 & p_{22} & \cdots & p_{2,n-1} & p_{2,n}\\
	\vdots & \vdots & \ddots & \vdots & \vdots\\
	0 & 0 & \cdots & p_{n-1,n-1} & p_{n-1,n}\\
	0 & 0 & \cdots & 0 & p_{n,n}
	\end{array} \right] \left[ \begin{array}{c}
	\beta_{1}\\
	\beta_{2}\\
	\vdots\\
	\beta_{n-1}\\
	\beta_{n}
	\end{array} \right]= \left[ \begin{array}{c}
	0\\
	0\\
	\vdots\\
	0\\
	b_{n}
	\end{array} \right]
	\end{equation}
	
	(2) Next, by the inductive method, the solution $ \overline{b}= \left[ \beta_{1}, \beta_{2}, \cdots, \beta_{n} \right]^{T}$ of the 
	triangular matrix equation \eqref{eq:e3d024} is proven as follows
	
	\begin{equation} \label{eq:e3d025}
	\beta_{n-k}= \frac{(-1)^{k}b_{n}}{(n-k-1)!k! \varDelta^{n-1}}, \qquad k=0,1, \ldots,n-1
	\end{equation}
	
	(2.1) Solving the last two equations in the triangular matrix equation \eqref{eq:e3d024}, we have
	\begin{align}
	\beta_{n} &	= \frac{b_{n}}{p_{n,n}}= \frac{b_{n}}{(n-1)! \varDelta^{n-1}} \\
	\beta_{n-1}	& = \frac{-p_{n-1,n}}{p_{n-1,n-1}} \beta_{n} = \frac{-(n-1)! \varDelta^{n-2}}{1!(n-2)! \varDelta^{n-2}} \times \frac{b_{n}}{(n-1)! \varDelta^{n-1}}= \frac{-b_{n}}{(n-2)! \varDelta^{n-1}}
	\end{align}
	that is, Eq. \eqref{eq:e3d025} holds for $k=0,1$.
	
	(2.2) It is assumed that Eq. \eqref{eq:e3d025} holds for $k<m \leq n-1$, that is , we have 
	\begin{equation} 
	\beta_{n-k}= \frac{(-1)^{k}b_{n}}{(n-k-1)!k! \varDelta^{n-1}}, \qquad k<m
	\end{equation}
	
	(2.3) Here, it will be to prove Eq. \eqref{eq:e3d025} holds for $k=m$, that is, it needs to prove the following equation hold.
	\begin{equation} 
	\beta_{n-m}= \frac{(-1)^{m}b_{n}}{(n-m-1)!m! \varDelta^{n-1}},
	\end{equation}
	
	According to the above assume, solving the $(n-m)$-th equation in Eq. \eqref{eq:e3d022} and then we have 
	\begin{align}
	\beta_{n-m}	& = \frac{-1}{p_{n-m,n-m}}
	\sum_{i=0}^{m-1}p_{n-m,n-i} \beta_{n-i}
	\notag \\ 
	& 
	= \frac{-1}{(n-m-1)! \varDelta^{n-m-1}} \times
	\sum_{i=0}^{m-1} \left[ \frac{(n-i-1)!}{(m-i)!} \varDelta^{n-m-1} \times \frac{(-1)^{i}b_{n}}{(n-i-1)!i! \varDelta^{n-1}} \right]
	\notag \\ 
	& 
	= \frac{-b_{n}}{(n-m-1)!m! \varDelta^{n-1}} \times 
	\sum_{i=0}^{m-1} \frac{(-1)^{i}m!}{(m-i)!i! } 
	\notag \\ 
	& 
	= \frac{-b_{n}}{(n-m-1)!m! \varDelta^{n-1}} \times \left[(1-1)^m-
	\frac{(-1)^{m}m!}{m!0! } \right]
	\notag \\ & 
	= \frac{(-1)^{m}b_{n}}{(n-m-1)!m! \varDelta^{n-1}}
	\end{align}
	Therefore, Eq. \eqref{eq:e3d025} for $k=m$ holds also.
	
	In summary, Eq. \eqref{eq:e3d025} is proven hold by the inductive method.
	
	(3) Based on the solution of the triangular matrix equation \eqref{eq:e3d024} and \textbf{ Theorem \ref{la:ab02} }, the volume of the zonotope generated by the matrix pair $(\overline{A},\overline{b})$ with the eigenvalues $ \left \{ \lambda_{i}= \lambda+(i-1) \varDelta,i= \overline{1,n} \right \} $ is 
	\begin{align}
	V( \varDelta) &	= \left| \det (P) \left[ \prod_{0<i<j \leq n} \frac{ \lambda_{j}- \lambda_{i}}{1- \lambda_{j} \lambda_{i}} \right] \left[ \prod_{i=1}^{n} \frac{ \beta_{i}}{1- \lambda_{i}} \right] \right|
	\notag \\ 
	&
	= \left( \prod_{i=1}^{n-1}i! \right) \varDelta^{n(n-1)/2} \left[ \prod_{0<i<j \leq n} \frac{(j-i) \varDelta}{1- \lambda_{j} \lambda_{i}} \right] \left| \left[ \prod_{i=1}^{n} \frac{(-1)^{(n-i)}b_{n}}{(i-1)!(n-i)! \varDelta^{n-1} \left(1- \lambda_{i} \right)} \right] \right|
	\notag \\ 
	&= \left( \prod_{i=1}^{n-1}i! \right) \varDelta^{n(n-1)/2} \frac{ \left[ \prod_{k=1}^{n}(n-k)! \right] \varDelta^{n(n-1)/2}}{ \prod_{0<i<j \leq n} \left(1- \lambda_{j} \lambda_{i} \right)} \left[ \prod_{i=1}^{n} \frac{1}{(i-1)!(n-i)! \left(1- \lambda_{i} \right)} \right] \frac{b_{n}^{n}}{ \varDelta^{n(n-1)}}
	\notag \\ 
	&
	=b_{n}^{n} \left[ \prod_{0<i<j \leq n} \frac{1}{1- \lambda_{j} \lambda_{i}} \right] \left[ \prod_{i=1}^{n} \frac{1}{1- \lambda_{i}} \right]
	\end{align}
	
	When $ \varDelta \rightarrow0$, the volume for the matrix pair $(\overline{A},\overline{b})$ is indeed the volume for the Jordan matrix pair $(A,b)$. Therefore, the volume for the Jordan matrix pair $(A,b)$ is as 
	\begin{align}
	\lim_{ \varDelta \rightarrow0}V( \varDelta)	& = \lim_{ \varDelta \rightarrow0}b_{n}^{n} \left[ \prod_{0<i<j \leq n} \frac{1}{1- \lambda_{j} \lambda_{i}} \right] \left[ \prod_{i=1}^{n} \frac{1}{1- \lambda_{i}} \right]
	\notag \\ 
	&
	= \frac{b_{n}^{n}}{ \left(1- \lambda \right)^{n} \left(1- \lambda^{2} \right)^{n(n-1)/2}}
	\end{align}
	Thus, the theorem has been proven. 

	Based on the theorem, the volume of the zonotope generated by Jordan matrix pair can be computed analytically and conveniently. And then, the volume of the zonotope generated by the general matrix pair $(A,B)$ can be computed as follows
	\begin{align}
	\lim_{ \varDelta \rightarrow0}V( \varDelta)	= \frac{
		\left \vert \det \left( P_J\right ) \left( q_n b_{n} \right) ^{n} \right \vert }{ \left(1- \lambda \right)^{n} \left(1- \lambda^{2} \right)^{n(n-1)/2}}
	\end{align}
	where the matrix $P_J$ is the Jordan transformation matrix and the vector $q_n$ is the only unit left eigenvector of the matrix $A$.
	
	\subsection{The volume computation of the infinite-time zonotope for the Jordan matrix $A$ with multiple Jordan blocks }
	
	If the matrix $A$ is a Jordan matrix with multiple Jordan blocks, the Jordan matrix pair $(A,b)$ can be denoted by
	\begin{align}
	(A,b)=\left( \textnormal{diag-block} \left\{ J_1, J_2, \dots, J_q \right\} ,
	\left [\beta_1^T, \beta_2^T, \dots, \beta_q^T \right]^T \right) \label {eq:as002}
	\end{align}
	where $q$ is the Jordan block number of the Jordan matrix $A$, the matrix block $J_i$ is a $m_i \times m_i$ Jordan block with the eigenvalue $\lambda_i$, the matrix block $\beta_i$ is $\left [b_{i,1}, b_{i,2}, \dots, b_{i,m_i} \right]^T$. And then, we have
	$$ \sum _{i=1} ^{q} m_i =n $$
	
	Similar to \textbf{ Theorem \ref{c3bt01}} for the Jordan matrix with only one Jordan block, a theorem about the volume computation of the infinite-time zonotope for the Jordan matrix $A$ with multiple Jordan blocks can be stated as follows.
	
	\begin{theorem} \label{c3bt02} When all eigenvalues of the matrix $A$ satisfy $ \lambda \in [0,1) $, the volume of the infinite-time zonotope generated by the Jordan matrix pair $(A,b)$ as Eq. \eqref {eq:as002} is computed as follows 
		\begin{equation} \label{eq:e3d0192}
		V = \left \vert \prod _{i=1}^{q-1} \prod _{j=i+1}^{q} \left( \frac{\lambda_i - \lambda_j}{1- \lambda_i \lambda_j } \right)^{m_i \times m_j} \right \vert
		\times	
		\left \vert \prod _{i=1}^{q}	\frac{b_{i,m_i}^{m_i}}{ \left(1- \lambda_i \right)^{m_i} \left(1- \lambda_i ^2 \right)^{m_i(m_i-1)/2} }
		\right \vert
		\end{equation}
		When matrix $A$ is a general matrix, the volume of the infinite-time zonotope generated by the matrix pair $(A,b)$ is computed as follows 
		\begin{equation} \label{eq:e3d01921}
		V = \left \vert \prod _{i=1}^{q-1} \prod _{j=i+1}^{q} \left( \frac{\lambda_i - \lambda_j}{1- \lambda_i \lambda_j } \right)^{m_i \times m_j} \right \vert
		\times	
		\left \vert \det \left (P_J \right) \prod _{i=1}^{q}	\frac{\left(q_ib\right)^{m_i}}{ \left(1- \lambda_i \right)^{m_i} \left(1- \lambda_i ^2 \right)^{m_i(m_i-1)/2} }
		\right \vert
		\end{equation}
		where the matrix $P_J$ is the Jordan transformation matrix and the vector $q_i$ is the only unit left eigenvector of the matrix $A$ for the eigenvalue $\lambda_i$.
	\end{theorem}	
	
	The above theorem can be proven as \textbf{ Theorem \ref{c3bt01}}.

	\section{Decoding the Volume of the Controllability Region}
	
	According to the computing equation \eqref{eq:e3d01921}, some factors described the shape and size of the controllability religion, that is,  the zonotope generated by the matrix pair $(A,B)$ with the repeated eigenvalues, are deconstructed as follows.
	\begin{align} 
	F_1 & = \left \vert \prod _{i=1}^{q-1} \prod _{j=i+1}^{q} \left( \frac{\lambda_i - \lambda_j}{1- \lambda_i \lambda_j } \right)^{m_i \times m_j} \right \vert
	\times	
	\left \vert \prod _{i=1}^{q}	\frac{1}{ \left(1- \lambda_i ^2 \right)^{m_i(m_i-1)/2} }
	\right \vert
	\label{eq:63} \\
	F_{2,i,j} 
	&= \left \{ 
	\begin{array}{ll}
	\frac{ \left \vert q_{i,j}b \right \vert}{ 1- \lambda_{i} }
	& j=m_i \\
	\frac{ \left \vert q_{i,j}b +F_{i,j+1} \right \vert}{ 1- \lambda_{i} } & j=m_i-1,m_i-2,\dots
	\end{array}
	\right. \;\; i=1,2,\dots,q
	\label{eq:64} \\
		F_{3,i} & =\left \vert q_{i,m_i}b\right \vert^{m_i}
	\end{align}
	where all $ q_{i,j}b$ are assumed as with the same sign. These 3 classes of factors are similar to the last section, called as the shape factor, the side length of the circumscribed rhombohedral, and the modal controllability. These factors can be describe the shape and size of the zonotope, the control ability of the system, and the eigenvalue evenness factor of the linear system.
	
	\subsection {The Zonotope Shape Factor and the Eigenvalue Evenness Factor of the Linear System}

	Similar to the analysis for the matrix $A$ with $n$ different eigenvalues in last section, by Eq. \eqref{eq:e3d01921}, we can see, 
	when some two eigenvalues of the two Jordan blocks of the system matrix $A$ are approximately equal, the minimum distance of the boundary of the zonotope to the original of the stat space will be approximately zero, and the zonotope will be flattened.
	Therefore, the distributions of all eigenvalues of the matrix $A$ are even, the ratio between the minimum and maximum distance of the boundary to original can be avoided as a small value and the zonotope, that is, the control region, will be avoided flattened. And then, the volume of the zonotope and the control ability will maintain a certain size.
	
	The factor $F_1$ deconstructed from the volume computing equation \eqref{eq:e3d01921} can be used to describe the uniformity the $n$ distance of the boundary to the original in the $n$ eigenvector. The bigger the value of the factor $F_1$, the bigger 
	the ratio between the minimum and maximum distance of the boundary is, and then the greater the volume of the zonotope.
	
	Otherwise, the factor $F_1$ can be used to describe the evenness of the eigenvalue distribution of the linear system $\Sigma(A,B)$. The bigger the value of the factor $F_1$, the bigger the controllable region of the system is, and the stronger the control ability of the systems is.
	
	\subsection {The circumscribed hypercube and circumscribed rhombohedral of the reachability region}
			
\begin{figure}[htbp]
		\centering
		\includegraphics[width=0.8\textwidth]{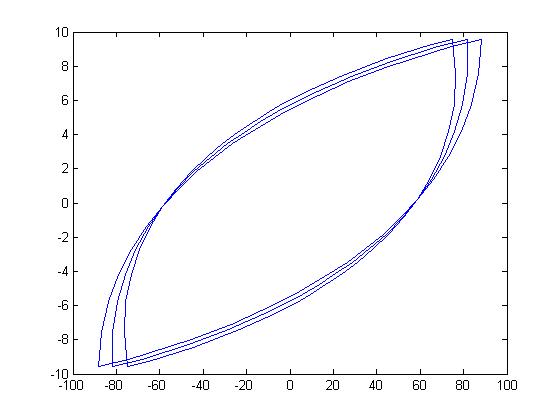} 
		\caption[c]{The 3 2-D zonotopes for Jordan matrix pairs $(A,b)$ with the different $b$ \label {fig:as03} }	
	\end{figure}

	Fig. \ref{fig:as03} shows the 2-dimensional zonotopes by the 3 Jordan matrix pairs $(A,b)$ as follows
	$$
	A=\left[ \begin{array}{cc}
	0.9 & 1 \\
	0 & 0.9
	\end{array} \right] \;\; 
	b=\left[ \begin{array}{c}
	0.7 \\ 1
	\end{array} \right],\;\left[ \begin{array}{c}
	0 \\ 1
	\end{array} \right] , \; \left[ \begin{array}{c}
	-0.7 \\ 1
	\end{array} \right] 
	$$
	
	By Fig. \ref{fig:as03}, we know, the factor $F_{2,i,j}$ is indeed the biggest distance of the boundary of the zonotope in the each eigenvector, that is, the $n$ side lengths of the circumscribed hypercube of the zonotope in the invariant eigen-space are $2F_{2,i,j},i=\overline{1,n}$. By the volume equation \eqref{eq:e3d01921}, the volume of the zonotope region can be represented as the production of the volume $\prod _{i=1} ^{n} F_{2,i,j}$ of the circumscribed hypercube and the shape factor $F_1$
	
By \textbf {Theorem, \ref{th3201}}, we know, the volume of the zonotope generated by the Jordan matrix pair $(A,b)$ has relation only to the last rows of the matrix blocks of the matrix $b$, corresponding to the each Jordan block of the matrix $A$, and has no relation to other rows.
But from Fig. \ref{fig:as03}, we know, these 'other' rows of matrix $b$ of Jordan matrix pair $(A,b)$ don't affect the size of the volume but maybe affect the shape of the corresponding zonotopes.
 
 \section {Conclusions}
 
 In this paper, the analytical volume computations of the zonotopes generated by the matrix pair with $n$ different or repeated real eigenvalues are discussed. By deconstructing the volume computing equations, 3 classes of the shape factors are constructed. These analytical volume and shape factors can describe accurately the zonotopes.
Because the control ability for LDT systems with the unit input variables (i.e. the input variable is with bounded value as 1) is directly related to the reachability region \cite{zhaomw202003}, based on these analytical expressions on the volume and shape factors, the control ability can be quantized conveniently. By choosing these analytical expressions as the objective function or constraint conditions, a novel optimizing problems and solving method for the control ability can be founded. Based on the optimization, not only the open-loop control ability, but also the some closed-loop control performance, such as the optimal time waste, robustness of the control strategy, etc, can be promoted, according to the conclusions in paper \cite{zhaomw202003} .
 
\bibliographystyle{model1b-num-names}
\bibliography{zzz}
\end{document}